\documentstyle[11pt,newpasp,twoside,epsf]{article}
\def\edcomment#1{\iffalse\marginpar{\raggedright\sl#1\/}\else\relax\fi}
\reversemarginpar

\begin{document}
\title{WFCAM, UKIDSS, and ${\mathrm z}=7$ quasars}
 \author{Steve Warren}
\affil{Astrophysics, ICSTM, Prince Consort Rd, London SW7 2BW, UK}
\author{Paul Hewett}
\affil{Institute of Astronomy, Madingley Rd, Cambridge, CB3 OHA, UK}

\begin{abstract}
We estimate the number of ${\mathrm z} \simeq 7$ quasars that will be
discovered in the Large Area Survey (LAS) element of the UKIRT
Infrared Deep Sky Survey (UKIDSS). The LAS will cover 4000 sq degs of the
northern sky to $K=18.4$, which is 3 mag. deeper than 2MASS. The Sloan
Digital Sky Survey has extended the quasar redshift limit to ${\mathrm
z}=6.3$. We demonstrate that to reach higher redshifts ${\mathrm
z}\sim7$, when Ly$\alpha$ has passed through the $z$ band,
combinations of standard broad-band filters such as $zJH$ and $zJK$
are ineffective. Instead the wavelength range between $z$ and $J$ must
be exploited. We introduce the $Y$ passband $0.97-1.07\mu{\mathrm m}$
for this purpose. High-redshift quasars up to redshift ${\mathrm
z}=7.2$ can be selected from a $iYJ$ or $zYJ$ two-colour diagram, as
bluer than L and T dwarfs.
\end{abstract}

\section{Motivation}
Over the past two years the Sloan Digital Sky Survey (SDSS) has had
considerable success in searches for quasars of very high redshift
${\mathrm z}>5$ (e.g. Schneider, these proceedings). Currently the
highest redshift recorded is ${\mathrm z}=6.3$ (Fan et
al., 2001a). Perhaps the most important motivation for searching for
sources at even higher redshifts is the possibility of detecting and
studying the epoch of reionisation. Analysis of the optical depth
blueward of the Ly$\alpha$ emission line in the spectrum of the
${\mathrm z}=6.3$ quasar indicates that we may be on the threshold of
this event (Becker et al., 2001). The detection of quasars of higher
redshift could probe the neutral IGM during or even before the epoch
of reionisation. Although the optical depth in Ly$\alpha$ is too high
to provide a useful measure of the neutral fraction in the IGM
(e.g. Madau, these proceedings), other species, observable with NGST,
might provide a measure. Also other probes of the IGM have been
suggested (e.g. Loeb and Rybicki, 2000).

\begin{figure}
\plotone{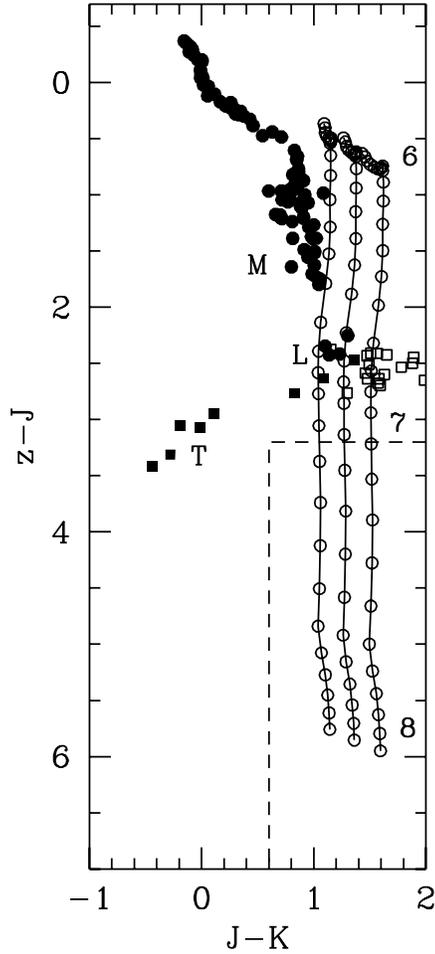}
\caption{Two-colour $zJK$ diagram showing colours of stars O to M
(filled circles), and L and T brown dwarfs, open and filled squares
respectively. All magnitudes are on the Vega system. The chains show
model quasar colours $5<{\mathrm z}<8$, $\Delta {\mathrm z}=0.1$, with
three different continuum slopes. This two-colour diagram is
ineffective for selecting quasars in the redshift range $6<{\mathrm
z}<7$, and other combinations of $zJHK$ are no better. Quasars
${\mathrm z}>7$ could be selected by identifying objects $z-J>3.2$,
i.e. in the box shown. But to reach $J=19$ requires a depth $z=22.2$,
whereas the SDSS only reaches $z=19.9$.}
\end{figure}

\begin{figure}
\plotone{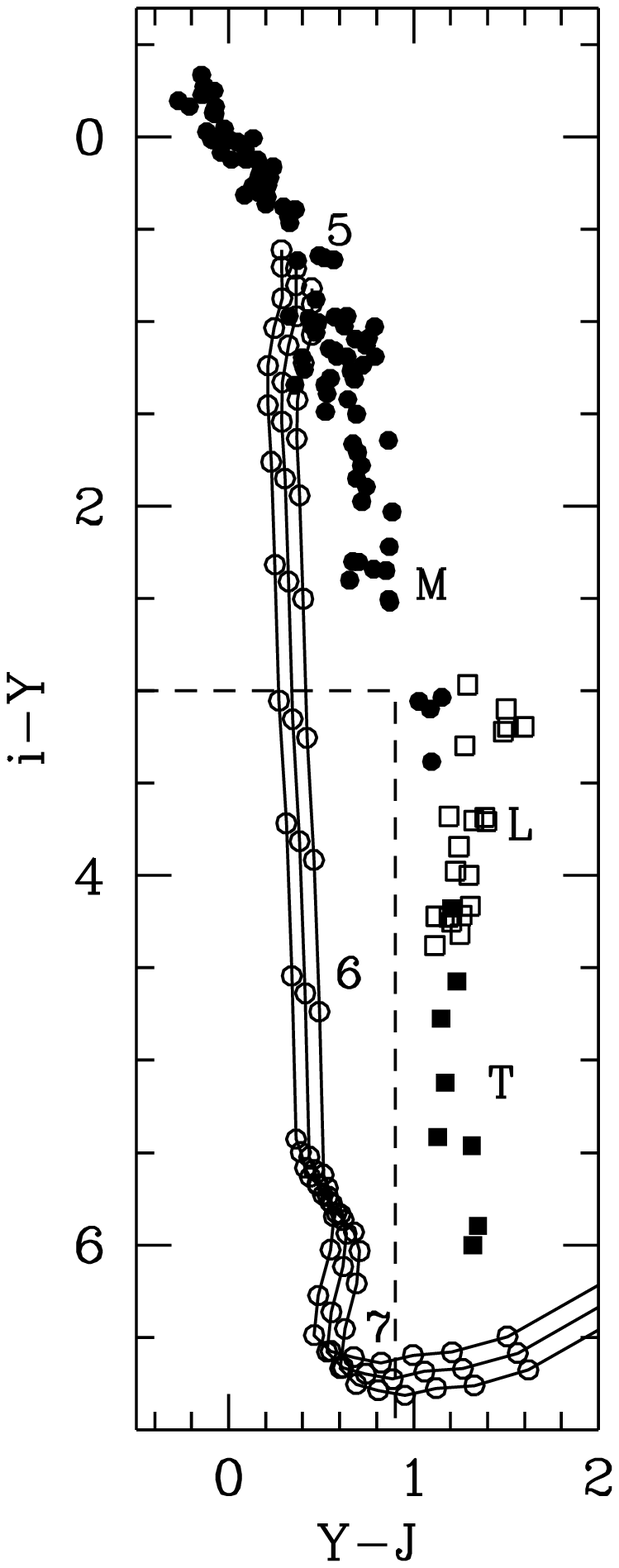}
\caption{Two-colour $iYJ$ diagram showing colours of stars O to M
(filled circles), and L and T brown dwarfs, open and filled squares
respectively. All magnitudes are on the Vega system. The chains show
model quasar colours $5<{\mathrm z}<8$, $\Delta {\mathrm z}=0.1$, with
three different continuum slopes. Any objects $i-Y>3$, in the
selection box shown, are candidate quasars in the redshift range
$5.8<{\mathrm z}<7.2$. Because the SDSS reaches $i=22.0$ this works to
a limit $Y=22.0-3.0=19.0$. In fact the UKIDSS LAS will reach $Y=20.5$
so deeper optical data will allow the discovery of fainter
high-redshift quasars.}

\end{figure}

\section{Higher redshifts than SDSS}
The multicolour technique for finding high-redshift quasars exploits
the break in the spectrum across Ly$\alpha$, caused by Ly$\alpha$
absorption. The highest redshift quasars in the SDSS are detected in
an $izJ$ two-colour diagram. A quasar at ${\mathrm z}\sim 6$ is very
red in $i-z$, but blue in $z-J$ relative to other sources with the
same $i-z$ colours i.e. L and T dwarfs. The passbands and depths of
the SDSS mean that the survey is unlikely to find any quasars beyond
${\mathrm z}=6.5$ because at these redshifts the quasars become too
faint to be detected in the $z$ band, due to Ly$\alpha$ absorption
(Fan et al. 2001a).

To reach higher redshifts it is natural to consider shifting the
filter set to longer wavelengths, and using the combination $zJH$ or
$zJK$. This does not work however, because the energy peak of cool
stars lies in the near-infrared, so that quasars are no longer bluer
than cool stars. This is illustrated in Fig. 1 which shows predicted
colours of quasars in the redshift range $5<{\mathrm z}<8$. The
quasars lie close to, or within the stellar locus, and do not emerge
until $z-J>3.2$. This requires prohibitively deep observations in the
$z$ band.

\section{The UKIDSS Large Area Survey}

The UKIRT Infrared Deep Sky Survey (UKIDSS, www.ukidss.org) is the
next generation near-infrared sky survey, the successor to 2MASS. The
instrument is WFCAM, which is equipped with $4\times2048$ arrays,
giving 0.21 sq. degs coverage per exposure. UKIDSS will commence in
2003 and comprises five elements. One of these, the Large Area Survey
(LAS), is a wide-field survey over 4000 sq. degs to $K=18.4$.  This
depth is three magnitudes deeper than 2MASS. The UKIDSS LAS will be
the true near-infrared counterpart to the SDSS.

The problem in finding ${\mathrm z}=7$ quasars is that they have
similar colours to brown dwarfs at near-infrared wavelengths. However
by introducing a passband between $z$ and $J$ we can separate out the
quasars. UKIDSS will use a filter called $Y$ covering the wavelength
range $0.97-1.07\mu{\mathrm m}$ (some observers call this wavelength
range $z(ir)$, but since there is no overlap with the SDSS $z$ band we
prefer a distinct name). Fig. 2 compares the colours of high-redshift
quasars $5<{\mathrm z}<8$, with the colours of stars and brown
dwarfs. The quasars are well separated from the L and T dwarfs in the redshift range $5.8<{\mathrm z}<7.2$.

UKIDSS will use the SDSS data for the short-wavelength band. Either
the $i$ or the $z$ band is equally effective, because the shorter
lever arm of the $z-Y$ colour (better for detecting the spectral
break) compensates for the greater depth of the $i$ band. The SDSS $i$
limit is 22.0 which means that using a selection limit $i-Y>3$, as
shown in Fig. 2, UKIDSS can detect all quasars in the redshift range
$5.8<{\mathrm z}<7.2$, and brighter than $Y=19.0$. Using the luminosity
function of Fan et al. (2001b) we expect to detect about 10 quasars
over this redshift range over 4000 sq. degs. These are the very
brightest high-redshift quasars, and the most useful for follow-up
spectroscopy. However larger numbers could be detected using deeper
optical imaging, such as the proposed CFH Legacy Survey
(www.cfht.hawaii.edu/Science/CFHLS/).

\end{document}